\documentclass{article}

\usepackage[final]{nips_dlps_2017}

\usepackage[utf8]{inputenc} 
\usepackage[T1]{fontenc}    
\usepackage{hyperref}       
\usepackage{url}            
\usepackage{booktabs}       
\usepackage{amsfonts}       
\usepackage{nicefrac}       
\usepackage{microtype}      

\usepackage{amsmath}
\usepackage{graphicx}
\usepackage{subfig}

\usepackage{natbib}
\title{Towards understanding feedback from supermassive black holes using convolutional neural networks}

\author{
  Stanislav~Fort\\
  Stanford University\\
  Stanford, CA 94305, USA \\
  \texttt{sfort1@stanford.edu} \\
}

\begin{document}

\maketitle

\begin{abstract}
Supermassive black holes at centers of clusters of galaxies strongly interact with their host environment via \textit{AGN feedback}. Key tracers of such activity are \textit{X-ray cavities} -- regions of lower X-ray brightness within the cluster. We present an automatic method for detecting, and characterizing X-ray cavities in noisy, low-resolution X-ray images. We simulate clusters of galaxies, insert cavities into them, and produce realistic low-quality images comparable to observations at high redshifts. We then train a custom-built convolutional neural network to generate pixel-wise analysis of presence of cavities in a cluster. A ResNet architecture is then used to decode radii of cavities from the pixel-wise predictions. We surpass the accuracy, stability, and speed of current visual inspection based methods on simulated data.
\end{abstract}

\section{Introduction}

Galaxy clusters are the largest gravitationally bound structures in the observable Universe. Supermassive black holes typically reside in their cores, forming so-called \textit{Active Galactic Nuclei} (AGNs) -- black holes actively feeding on the cluster material. AGNs are able to affect the clusters and by extension the Universe at large via \textit{AGN feedback}. Understanding the mechanism, and its changes over the cosmic history is of great importance for astrophysics and cosmology.~\citep{HL}  

A key tracer of the \textit{AGN feedback} is the presence of \textit{X-ray cavities}. X-ray cavities look like regions of lower X-ray brightness in the cluster. Physically, they are volumes of gas that has been inserted into the intracluster medium via a black hole jet during periods of high accretion activity. Detecting and characterizing X-ray cavities at different distances, and therefore times, is crucial for understanding the evolution of the Universe.~\citep{HL}  

We are primarily concerned with clusters of galaxies at high redshift, which correspond to large distances. In such a regime, the resolution of images observed by space-based X-ray telescopes is low. Due to Poisson noise and low brightness, a typical pixel has zero photons incident on it. The images therefore look very noisy, and are hard to analyze in their raw form. The largest analysis of X-ray cavities in galaxy clusters so far was presented in \citet{HL}. Their detection method, however, relied on visual inspection. Upcoming space-based surveys will observe a large number of such objects, and therefore an automatic and principled approach to their analysis will be needed.

In this paper, we present a promising step towards automatic discovery and detection of X-ray cavities in high redshift clusters of galaxies. Using simulated data, we built and trained a convolutional neural network that makes a pixel-wise prediction of cavity presence in the image, surpassing human accuracy. We then feed the prediction in a ResNet-based~\citep{resnet} architecture that is able to estimate the radii of the cavities present. We recover the radii of inserted cavities with high accuracy on realistic simulated data.

The paper is structured as follows: We discuss cluster simulations in Section~\ref{sec:sim}. Our architecture is presented in Section~\ref{sec:arch}. Our experiments and their results are discussed in Section~\ref{sec:exp}. We conclude with Section~\ref{sec:conclusion}.

\section{Simulations}
\label{sec:sim}
Due to the inherent difficulty of detecting X-ray cavities in high-redshift clusters, a reliable training dataset is not available. Analyses of nearby, high-resolution, and high signal-to-noise clusters do exist \citep{HL}, however, they do not generalize well to large distances, and most likely contain a large number of false-positives. We therefore use simulations to generate our data. Employing physical insights from existing observations, we were able to construct a simple probabilistic model of a hydrostatically equilibrated cluster of galaxies. 

For each cluster, we randomly assign its brightness, radius, and position with respect to the center of the image. We used the $\beta$-profile \citep{beta} for its shape. We then add cavities of random radii, depths, and positions. This constitutes a simple probabilistic model of a cluster. Appreciating that, due to physical reasons, X-ray cavities often come in pairs, and that the predominant number of clusters will have either zero or one pair, we focused on such cases during training. However, our results do not specifically depend on any of these choices, and we are planning to extend them to more general scenarios.

A typical X-ray observation of a cluster of galaxies at high redshift will be very noisy. Examples of our simulated clusters are shown in Figure~\ref{fig:diagram1}. Notice that the actual input of our neural network is column (d) -- the raw photon counts. Typically, astronomers would visually inspect blurred raw images of clusters for cavities, such as shown in column (e). Due to the low photon counts, many cavity-like regions of lower brightness are visible there, presenting false positives that visual analysis struggles to deal with. Our network, however, seems to have learned to ignore them.
	\begin{figure}[th]
		\begin{center}
		\begin{tabular}{ccccc}
			{\includegraphics[width = 0.17\linewidth]{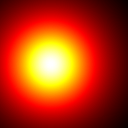}} &
			{\includegraphics[width = 0.17\linewidth]{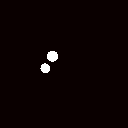}} &
			{\includegraphics[width = 0.17\linewidth]{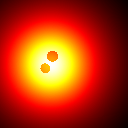}} &
			{\includegraphics[width = 0.17\linewidth]{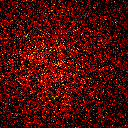}} &
			{\includegraphics[width = 0.17\linewidth]{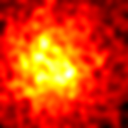}} \\
			{\includegraphics[width = 0.17\linewidth]{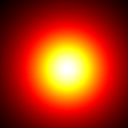}} &
			{\includegraphics[width = 0.17\linewidth]{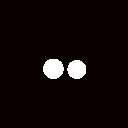}} &
			{\includegraphics[width = 0.17\linewidth]{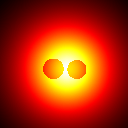}} &
			{\includegraphics[width = 0.17\linewidth]{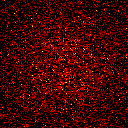}} &
			{\includegraphics[width = 0.17\linewidth]{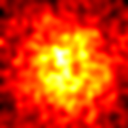}} \\  
			(a) Clean cluster & (b) Cavities mask & (c) Realistic cluster  & (d) Photon counts & (e) Blurred \\[6pt]
		\end{tabular}
		\caption{Examples of simulated data. Column (a) shows a simulated cluster of galaxies. Column (b) shows where X-ray cavities are inserted, and column (c) represents the cluster with the insertion. (d) is a realistic image of (c), as would be observed by space-based X-ray telescopes at high redshifts, and constitutes the \textit{input of our neural network}. The noise is due to Poisson statistics of photon counts. (e) shows a blurred version of (d) that would often be used for visual inspection by astronomers.}
		\label{fig:sim}
		\end{center}
	\end{figure}

\section{Methods}
\label{sec:arch}
\begin{figure}[th]
	\begin{center}
		\includegraphics[width=1.0\linewidth]{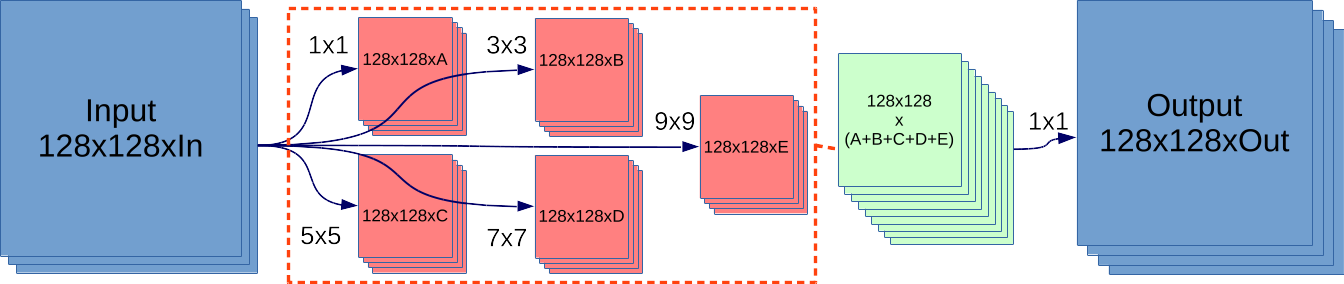}
	\end{center}
	\caption{A diagram of the encoder block. The image shows a diagram of one of the encoder blocks. An input is convolved with $A$ filters of sizes $1\times1$, $B$ $3\times3$, $C$ $5\times5$, $D$ $7\times7$, and $E$ $9\times9$ filters. The results of the convolutions are concatenated, and a $1 \times 1$ convolution is used on them to get the desired number of output channels. We found that this architecture works better than many off-the-shelf methods for mapping noisy images to pixel-wise cavity predictions.}
	\label{fig:diagram1}
\end{figure}

\begin{figure}[th]
	\begin{center}
		\includegraphics[width=1.0\linewidth]{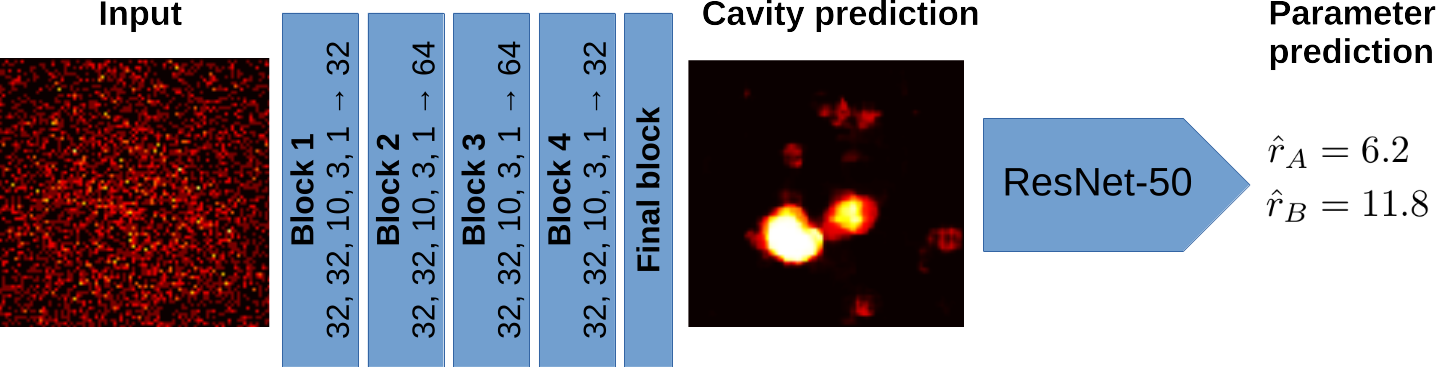}
	\end{center}
	\caption{A diagram of the architecture used. A realistic input image is fed into a sequence of 4 convolutional blocks (shown in detail in Figure~\ref{fig:diagram1}) and the final block, that contains 8 $8\times8$, 4 $16\times16$, and 2 $32\times32$ filters. The resulting pixel-wise prediction of cavity presence is then fed into a ResNet-50 \citep{resnet} architecture to extract cavity radii. Batch norms and ReLUs are omitted for clarity.}
	\label{fig:architecture}
\end{figure}

We develop a method that maps a noisy, low-resolution image of a galaxy cluster into an image of the same resolution, where each pixel value represents the confidence that the particular pixel belongs to a cavity. To do this, we designed a custom convolutional neural network architecture composed of convolutional blocks stacked in a sequence. The details of a block are shown in Figure~\ref{fig:diagram1}, and the whole architecture is presented in Figure~\ref{fig:architecture}. The block is designed such that it can incorporated information on different scales. Each convolutional block is followed by batch normalization and \verb|ReLU|, which we omitted in Figure~\ref{fig:architecture} for clarity. Our encoder architecture performed better than off-the-shelf methods for image segmentation we explored.

We calculate a cross-entropy loss of the predicted cavity mask with respect to the ground truth cavity mask known from simulation. The predicted cavity mask is then fed into ResNet-50 \citep{resnet}, and cavity radii are estimated. At the present stage, we limited the number of estimated radii to either zero or two, but our method allows for an arbitrary number in general. We use the squared distance between the target radii and predicted radii as loss, and only train the ResNet-50 part of our architecture using it.
\section{Experiments}
\label{sec:exp}
We employ online data generation during training, using a probabilistic model of a cluster of galaxies based on physically realistic parameters. This acts as a regularization for our network and prevents overfitting. An important part of the training set is the possibility that there are no cavities in a given cluster. We generate such examples $50 \%$ of the time to force the network to truly learn to detect rather than blindly guess that there are cavities near the center of the cluster. We trained our architecture for $40$ epochs with $9,000$ clusters in each at learning rates $10^{-3}$ and then $0.3 \times 10^{-3}$ using the Adam optimizer. Each cluster came with $3$ different noise realizations. We used 4 Nvidia Titan GPUs during training.

\begin{figure}[th]
		\begin{center}
			\begin{tabular}{cccc}
				{\includegraphics[width = 0.17\linewidth]{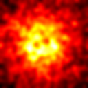}} &
				{\includegraphics[width = 0.17\linewidth]{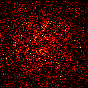}} &
				{\includegraphics[width = 0.17\linewidth]{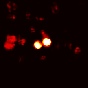}} &
				{\includegraphics[width = 0.17\linewidth]{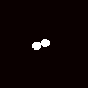}} \\
				{\includegraphics[width = 0.17\linewidth]{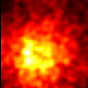}} &
				{\includegraphics[width = 0.17\linewidth]{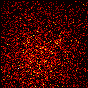}} &
				{\includegraphics[width = 0.17\linewidth]{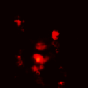}} &
				{\includegraphics[width = 0.17\linewidth]{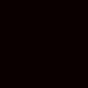}} \\
				(a) Blurred & (b) Input & (c) Prediction  & (d) Ground truth  \\[4pt]
			\end{tabular}
			\caption{Examples of test set results. Column (a) shows a blurred version of the input data (b), illustrating the difficulty of reliably detecting cavities by visual inspection. Column (c) is the prediction of our architecture and column (d) the ground truth. The match between the prediction and truth is very good, even in the case of no cavity being present (second row). These results are typical for our method.}
			\label{fig:exp}
		\end{center}
	\end{figure}

\begin{figure}[th]
	\begin{center}
		\begin{minipage}{.5\linewidth}
			\centering
			\includegraphics[width=1.0\linewidth]{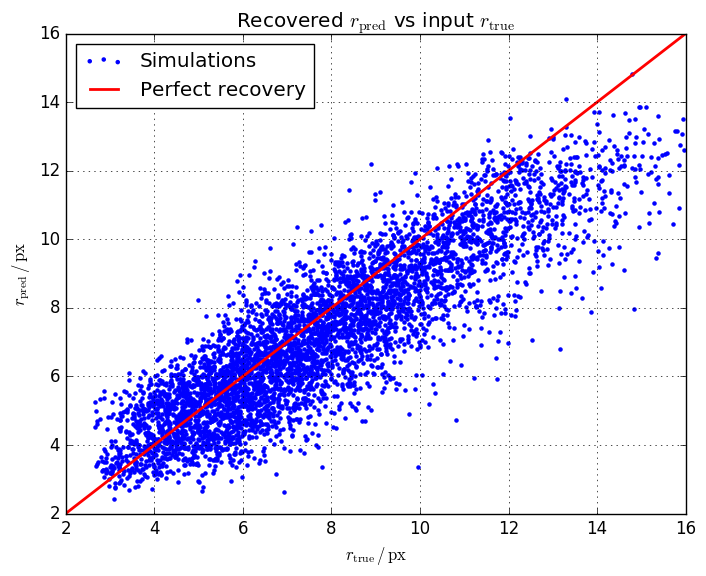}
		\end{minipage}%
		\begin{minipage}{.5\linewidth}
			\centering
			\includegraphics[width=1.0\linewidth]{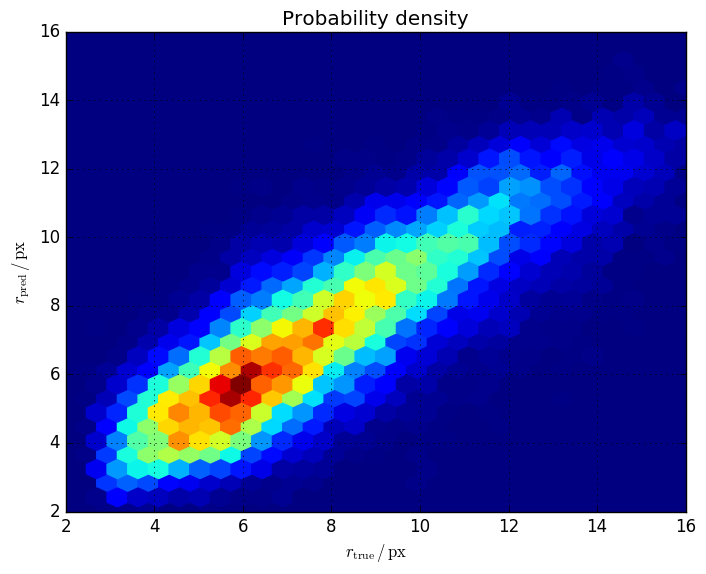}
		\end{minipage}
		\caption{A comparison of inserted and recovered radii of cavities. The $x$-axis shows the radius of a cavity inserted into a simulated cluster, and the $y$-axis corresponds to the value inferred by our architecture from its noisy image. A clear linear relationship between $r_{\mathrm{true}}$ and $r_{\mathrm{pred}}$ is visible, as also demonstrated by the probability density plot on the right-hand side.}
		\label{fig:rr}
	\end{center}
\end{figure}

Representative results of our experiments are shown in Figure~\ref{fig:exp}. In particular, we show a blurred version of the input to appreciate the difficulty of detecting cavities reliably by visual inspection alone, as used to be done by astronomers so far. We show that the we are able to reliably recover radii of inserted cavities using our architecture in Figure~\ref{fig:rr}. This is a promising result towards applying our method to real data.
\section{Conclusion}
\label{sec:conclusion}
We developed a convolutional neural network based approach to detection and characterization of X-ray cavities in low-resolution, noisy images of clusters of galaxies. Training and testing on realistic simulated data, we show that our method is robust to different kinds of noise, and that we are able to accurately and consistently predict pixel-wise cavity positions and shapes, as well as reliably extract their radii. This gives us hope towards applying our method to real observations. With the upcoming large-scale X-ray cluster surveys, an automatic method for cavity detection and analysis will be needed. In particular, our advantage over visual inspection is reliability, consistency, and at least $100$-fold speedup. We are also able to analyze the raw noisy images, and can therefore utilize all information available. In the future, we will verify our method on real images of X-ray clusters and compare our results to \citep{HL}.

\subsubsection*{Acknowledgments}
We would like to thank Rebecca Canning, Steve Allen, and Yihui Quek (all from Stanford University) for useful discussions. Some of the computing for this project was performed on the Sherlock cluster. We would like to thank Stanford University and the Stanford Research Computing Center for providing computational resources and support that contributed to these research results.
\bibliography{cavities_lib}

\end{document}